\newcommand{\tsecompldate}{Friday, 6th October, 2000}

\documentclass[12pt,a4paper]{article}
\newcommand{\abstracts}[1]{\begin{abstract} #1 \end{abstract}}
\newcommand{\Journal}[4]{#1 {\bf #2} (#4) #3}

\usepackage{amsfonts}

\newcommand{\tpretitle}[1]{}

\newcommand{\tprenote}[1]{\footnote{#1}}

\newcommand{\tnote}[1]{}
\newcommand{\tcomment}[1]{}

\newcommand{\href}[2]{{#2}}
\newcommand{\eprint}[1]{{\tt #1}}

\newcommand{\tsedevelop}[1]{{}}

\renewcommand{\href}[2]{{#2}{}}
\renewcommand{\eprint}[1]{\href{http://xxx.soton.ac.uk/abs/#1}{{\tt #1}}}

\newcommand{\half}{\frac{1}{2}}
\newcommand{\beq}{\begin{equation}}
\newcommand{\eeq}{\end{equation}}
\newcommand{\nnel}{\nonumber \\ {}}

\newcommand{\Real}{\mathbb{R}}

\newcommand{\Edd}{E''}

\newcommand{\Bret}{B_{\rm ret}}
\newcommand{\Seff}{S_{\rm eff}}
\newcommand{\Sefft}{S_{\rm eff}^{(2)}}
\newcommand{\Steff}{S^{(2)}_{\rm eff}{}}

\newcommand{\Vteff}{V^{(2)}_{\rm eff}}

\newcommand{\dslash}{{d\kern-.50em{/}}}

\newcommand{\intdtk}{\int \dslash^3\veck  \;}
\newcommand{\intdtp}{\int \dslash^3\vecp  \;}

\newcommand{\intdfx}{\int d^4x \;}

\newcommand{\veck}{\vec{k}}
\newcommand{\vecp}{\vec{p}}
\newcommand{\vecx}{\vec{x}}

\newcommand{\calL}{{\cal L}}

\newcommand{\phibar}{\bar{\phi}}
\newcommand{\phidel}{{\phi}_{\delta}}

\begin{document}

\renewcommand{\thefootnote}{\fnsymbol{footnote}}

\begin{flushright}
Imperial/TP/00-01/001 \\
{\tt hep-ph/0010138} \\
\tsecompldate \\
\end{flushright}

\begin{center}
{\Large\bf Time Dependent Effective Actions at Finite
Temperature\footnote{Poster given at SEWM 2000 Marseille, 15th
June 2000. Parts of this work done in collaboration with
M.Asprouli, V.\ Galan Gonzalez and R.Rivers, and a
non-relativistic version with D.Steer.  This preprint version
contains additional footnotes and appendices.}}
\\ \vskip 1cm
{\large
\href{http://theory.ph.ic.ac.uk/links/time}{T.S.Evans}\footnote{E-mail:
\href{mailto://T.Evans@ic.ac.uk}{\tt T.Evans@ic.ac.uk}, WWW:
\href{http://theory.ph.ic.ac.uk/links/time}{\tt
http://theory.ph.ic.ac.uk/$\sim$time}}}
\\[1cm]
\href{http://theory.ph.ic.ac.uk/}{Theoretical Physics},
\href{http://www.ph.ic.ac.uk/}{Blackett Laboratory},
\href{http://www.ic.ac.uk/}{Imperial College},\\
Prince Consort Road, London SW7 2BW  U.K.
\end{center}

\abstracts{I study derivative expansions of effective actions at
finite temperature, illustrating how the standard methods are
badly defined at finite temperature.  I then show that by setting
up the initial conditions at a finite time, these problems are
solved.}

\renewcommand{\thefootnote}{\arabic{footnote}}
\setcounter{footnote}{0}


\section{Standard approach}

\subsection{Zero temperature}

I will work with a simple model of two real fields $\phi$ and
$\eta$\tprenote{This is a toy model of QED, where the $\eta$ field
plays the role of the electrons, and I am looking for an effective
theory for the photon, here played by the $\phi$ field.  In the
case of QED this would then be the Euler-Heisenberg effective
action.}
\begin{equation}
\calL [\phi,\eta] = \half \eta \Delta^{-1}{\eta}  - \half g \phi
\eta^2 + \calL_0 [\phi] ,
\label{Lscalar}
\end{equation}
and integrating out the {$\eta$} field gives
\begin{eqnarray}
 Z & = & \int D\phi {D\eta} \;
 e^{i \int d^4x \; \calL[\phi,   {\eta}] }
 = \int D\phi \; e^{i \int d^4x \;
 \calL_0[\phi] } \; e^{i {\Seff[\phi]} } .
\end{eqnarray}
Here the integration is exact  and the effective action for
$\phi$ is the classical part ${\cal L}_0[\phi]$ plus
\begin{eqnarray}
\Seff[\phi] &=& \frac{i}{2} {\rm Tr} \left\{ \ln \left[ 1 - g
\phi(x) \Delta(x,y) \right] \right\} .
\end{eqnarray}
This $\Seff [\phi]$ contains {\em all} $\eta$ fluctuations, both
quantum and statistical, even though it can only be used to
describe the behaviour of $\phi$. The problem is that $\Seff
[\phi]$ is still too complicated e.g.\ its non-local in $\phi(x)$.
The solution is to use further approximations, firstly expanding
the logarithm, and then to perform a derivative expansion.

Expanding the  $\ln$ gives $\Seff[\phi] = \Seff^{(1)} + {\Sefft} +
\ldots$ where
\begin{eqnarray}
 \Sefft &:=&
 -\frac{ig^2}{4} \int d^4x \int d^4y  \left\{ \phi(x) \Delta(x,y)
 \phi(y)  \Delta(y,x) \right\} \label{Sefftdef}
\end{eqnarray}
I will focus on {$\Sefft$} as this is quadratic in $\phi$, so it
contains important effective mass and kinetic terms for $\phi$ and
it is also the first term which shows all the features of the
problem.  The truncation of the {$\ln$} expansion is valid for
{weak coupling} and/or {weak fields}.

This term {$\Sefft$} is  however non-local in $\phi$ so we expand
$\phi(y)$ in terms of the field at $x$ to get infinite number of
local terms
\begin{eqnarray}
{\phi(y)} &=&    {\phi(x)} + (y-x)^\mu \partial_\mu    {\phi(x)} +
 \ldots
 = e^{ i(y-x)^\mu P_\mu }    {\phi(x)} , \; \; \;
 P_\mu = -i \partial_\mu
\end{eqnarray}
Truncating this derivative expansion gives a tractable if
approximate effective action with finite number of local terms.
This truncation is valid for fields varying slowly in time and
space.

\subsection{Finite Temperature}

It is easiest in this case to work in time not energy variables,
so the integrals over real times $t$ are replaced by integrals
along a directed path $C$ in the complex time plane. As this is a
dynamical problem, I will use the CTP (Closed Time Path)
approach\cite{LvW} where $C$ has three sections. The first, $C_1$
runs along the real axis from $t_i$ to $t_f$. $C_2$ is $C_1$ but
running in the opposite direction.  Finally there is the vertical
section, $C_3$ running from $t_i$ to $t_i-i\beta$.

The usual assumptions made are $t_i \rightarrow -\infty$, $t_f
\rightarrow +\infty$, and an equilibrium background field $\eta$.
For the latter the propagator is
\begin{eqnarray}
 \Delta_c (\tau, \tau'; \veck) &=&
 \frac{-i}{2\pi}\int_{-\infty}^{\infty} dE \; e^{-iE(\tau - \tau')}
 \left[ \theta_C(\tau ,\tau') + N(E) \right] \rho(E,\veck).
 \label{prop}
\end{eqnarray}
$\theta_C$ is the contour theta function\cite{LvW} -
$\theta_c(\tau, \tau')$ is $1$ ($0$) if $\tau$ ($\tau'$) is
further along $C$ than $\tau'$ ($\tau$). $N(E)$ is the
Bose-Einstein distribution $N(E) := [ \exp \{ \beta E \} -1]^{-1}$
and $\rho$ is the spectral function, which for a free relativistic
scalar field is of the form $\rho(E,k) = \pi (\omega(k))^{-1} [
\delta(E-\omega(k)) - \delta(E+\omega(k)) ]$. I will choose
$\omega(k)= (\veck^2+m^2)^{1/2}$ for the $\eta$ field dispersion
relation but the arguments below work with an arbitrary form.

The problem can be seen in the energy representation as the
derivative expansion of {$\Sefft$} can be rewritten in terms of
the small four-momentum expansion of the bubble diagram $B(P)$
\begin{eqnarray}
\Sefft &=& \intdfx \left[ \right. B(P=0) \; \phi(x)^2 +
 \left(\frac{\partial B(P)}{\partial E} \right)_{P=0}
 \phi(x) \frac{\partial}{\partial \tau}\phi(x)
\nnel &&
 \; \; \; \; \; \; \; \;
 \left.
 + \half  \frac{\partial^2 B(0)}{\partial E^2}
 \phi(x) \frac{\partial^2}{\partial \tau^2}\phi(x)
 + \ldots \right] ,
 \\
 B(P) &:=& \int_\beta \dslash^4 K \; \Delta(K) \Delta(K+P)  .
\end{eqnarray}
However, the equilibrium $B$ does {\em not} have a unique momentum
expansion at any $T>0$, something known in the non-relativistic
context at least since the work of Abrahams and Tsuneto in
1966\cite{AT} and for relativistic fields by Fujimoto in
1984\cite{Fu,TSEze}. There is therefore a serious problem when
trying to take the standard analysis to non-zero temperatures.


\section{A new approach}

The $T>0$ analysis above followed the usual approach without too
much attention to the physics of the problem.  I will therefore
try to work through it more carefully.  I produce a slightly
different result, one which will produce one solution to all these
problems. I will work with time rather than energy variables, and
to assume that the $\eta$ field starts and remains in equilibrium
for all time, so that the propagator $\Delta$ is unchanged. The
key lies in the values given to the c-number valued $\phi$ field
in the effective action.

First for times lying on $C_3$, the field $\phi$ represents
contributions coming from the initial density matrix $\rho$.
Expressing the density matrix in the Heisenberg picture, $\rho_H$,
in terms of interaction picture operators gives
\begin{equation}\label{dmatrixint}
  \widehat{\rho}_H := e^{ -\beta \widehat{H}_{\rm free,init} } .
  \widehat{U}_{\rm init}  , \; \; \;
  \widehat{U}_{\rm init}
  := T_C \exp \{ - i \int_{C_3} d\tau  \widehat{H}_{\rm int,init}  \}.
 \label{Udef}
\end{equation}
where the Hamiltonian at the initial time and in the interaction
picture is split as $\widehat{H}_{\rm init} = \widehat{H}_{\rm
free,init} + \widehat{H}_{\rm int,init}$. The factor of
$\widehat{U}_{\rm init}$ then generates Feynman diagrams with
vertices running over times equal to $t_i$ plus a variable pure
imaginary time component. It is therefore crucial {\em not} to use
the derivative expansion to express the fields coming from
$\widehat{U}_{\rm init}$ in terms of field values at later real
times. Instead for $\tau \in C_3$ I set $\phi(\tau,\veck) =
\phi_i(\veck)$, a time independent constant initial field, to be
specified as part of the initial conditions.

This is to be contrasted with the real-time fields $\phi(\tau)$
for $\tau \in C_1 + C_2$.  These are not associated with the
initial density matrix so I enforce no special initial conditions
on these fields. However, I will use the derivative expansion on
these real-time fields.  Also I will not force the field to have
the same value at the same real-time value if it is representing
points on $C_1$ and $C_2$ unless we wish to refer to classical
field configurations.

For simplicity an initial condition $\phi(t_i,\vecp)=0$ is assumed
here.\tnote{Further I will continue to focus on the behaviour of
the real time self-energy, rather than terms cubic and higher in
the field $\phi$.} Under this simplification, in this simple model
I have then
\begin{eqnarray}\label{s2eff3}
\Steff [\phi] &=&
- \half \sum_{a,b=1}^{2} \int_{C_a} d\tau \;
  \intdtp \phi_a(\tau,\vecp) B_{ab}\; \phi_b(\tau,-\vecp) ,
\end{eqnarray}
where $B$ is a bubble diagram given by
\begin{eqnarray}
 B_{ab}(\tau,\tau_i,\vecp,\Edd) &=& \frac{ig^2}{2}
 \int_{C_b} d\tau' \; \int \frac{d^3 \veck}{(2\pi)^3}
 \Delta_{ab}(\tau,\tau',\veck) \Delta_{ba}(\tau',\tau,\vecp+\veck)
 \nnel
 && \; \; \; \; \; \; \; \times \;
 \exp \{ -i\Edd(\tau'-\tau)\} ,
 \label{Gdef}
 \\
 \phi_a(\tau) := \phi(\tau), &&
 \Delta_{ab}(\tau,\tau',\veck) := \Delta_c(\tau,\tau', \veck)
 \; \; \; \; \tau \in C_a, \;  \tau' \in C_b  .
\end{eqnarray}
Note that $-i\Edd \equiv  \partial / \partial t$ ($t \in \Real $)
and $B_{ab} (a,b=1,2)$ are {\em operators} acting on $\phi_b
(b=1,2)$ only (not $b=3$ due to initial conditions). $B$ satisfies
the algebraic identities $B_{11}+B_{12} + B_{21} + B_{22} = 0$ and
$B_{13}+B_{23} = 0$.

As the only physical solution for field expectation values at real
times is $\phi_1(t)=\phi_2(t)$, it is useful to rewrite this as
\begin{equation}\label{s2eff4}
\Steff [\phi] = -\half \int_{C_1} dt \intdtp
 \begin{array}{c}
 ( \phibar(t,\vecp)  ,  \phidel(t,\vecp))
 \\
 \mbox{ }
 \end{array}
 \left( \begin{array}{cc}
 0 & B_{\rm adv}
 \\
 \Bret & B_{\rm fluc}
 \end{array}
 \right)
 \left( \begin{array}{c}
 \phibar(t,-\vecp)
 \\
 \phidel(t,-\vecp)
 \end{array}
 \right)  ,
\end{equation}
where $\phibar= (\phi_1+\phi_2 )/\surd 2$, $\phidel=
(\phi_1-\phi_2 )/ \surd 2$ at any one time and three-momentum.
$\Bret = B_{11} + B_{12}  = -B_{22} - B_{21}$ is one key object of
interest, as it is this term which appears in the equations of
motion so I will focus on this term. The result is
\begin{eqnarray}
{\Bret}
 &=& \frac{g^2}{8} \intdtk \sum_{ s_0,s_1=\pm 1} \frac{ s_0 s_1}{
 \omega \Omega}
 \left(1+N(s_0\omega) + N(s_1\Omega)\right)
 {\frac{( 1 - e^{i(t_i-t) {A}})}{{A}}}
 \\
 {A} &=& {E +s_0 \omega +s_1 \Omega},
 \; \; \; \omega=\omega(k), \; \; \; \Omega= \omega(k+p)  .
\end{eqnarray}
The Landau damping terms come from the $s_0=-s_1$, $\omega-\Omega$
factors.  In the limit of interest for  derivative expansion $E,p
\rightarrow 0$ so that $\Omega \rightarrow \omega$ and thus these
are dangerous as the denominator $A \rightarrow 0$.   In my case
though I have a crucial $t_i$ dependent term in the numerator
which ensures my numerator also goes to zero in this limit and my
expression is well behaved. Thus my $\Bret$ has {\em unique}
derivative series as its analytic about $E,p \rightarrow 0$, $A =
0$.

I do {\em not} get the traditional equilibrium result for $\Bret$
and the difference is the unusual factor $\exp \{ i (t_i-t) A \}$.
In equilibrium calculations using pure imaginary time methods, the
real external energy $E$ is replaced by $E-i\epsilon$ ($\epsilon$
is a real positive infinitesimal) during analytic
continuation\cite{LvW}. Then one takes the $t_i \rightarrow
-\infty$ limit.  This $t_i$ dependent term is then removed but the
integrand is then singular in the $\epsilon \gg E,p \rightarrow 0$
limit.

There are alternative solutions which work by keeping $A \neq 0$
in the zero momentum limit e.g. including thermalisation
rates/complex dispersion relations for the $\eta$
field\cite{BGR,AL} or keeping the masses in the two propagators
different\cite{AVBD}. However it is achieved, what is happening in
all cases is that a long time scale is being introduced and this
sets a regulator for this long time and long distance (small $E$
and $\vecp$) problem. In my case, it is more obvious as it is an
actual physical time $t_i$ rather than an energy parameter which
is performing the regulation.\tprenote{Of course another side
effect is that my results depend explicitly on $t_i$, the time at
which the initial conditions were setup. This should be expected
as the problem is {\em not} an equilibrium one even if some of my
fields are to an approximation in equilibrium. After all, looking
at small frequency perturbations means that you are implicitly
probing long time scales which will inevitably probe the time at
which the system was initialised.}

There are several conclusions to be drawn from this work. First we
have shown how to obtain a unique expansion for {\em weak, slowly
varying} fields in a heat bath with the $\omega-\Omega$ Landau
damping terms giving the dominant contribution. In particular,
when contributions from the vertical part are included this
analysis does show how the usual free energy results are the
lowest term in a consistent derivative expansion of an effective
action, as found at zero temperature.\tprenote{With the vertical
contributions (none-zero if $\phi \neq 0$ initially) and a
constant $\phi$ field, I get usual results for free energy.} This
approach also solves a lack of analyticity problem inherent in
linear response calculations.  Though the analysis has been
presented for a simple relativistic model, the principles are
universal, e.g.\ they work for a BCS superconductor\cite{ES}.

The biggest remaining problem is that there are time dependent
U.V. divergences ($\sim \ln (t-t_i)$ in the equations of motion),
presumably reflecting the fact that at $t_i = t$ we have no new
time scale in the game to set the scale for the low energy/long
time behaviour of my fields.\tprenote{This latter problem will be
avoided if other long time scales are included, such as damping in
propagators\cite{BGR,AL}.  Perhaps one also needs to ensure that
the initial conditions are set more consistently.}


\section*{Acknowledgments}
Parts of this work done in collaboration with M.Asprouli, V.\
Galan Gonzalez and R.Rivers.  I also wish to thank I.Aitchison,
I.Lawrie, I.Moss, and A.Schakel for useful discussions.


\section{Appendix}

\subsection{CTP curve}

\typeout{figure: CTP3 }
\begin{center}
\setlength{\unitlength}{5pt}
\begin{picture}(80,35)(-5,48)
\put(45,69){\makebox(0,0)[lb]{\large $t_f$}}
\put(25,70){\makebox(0,0)[lb]{\large $t$}}
\put(6,70){\makebox(0,0)[lb]{\large $t_i$}}
\put(7,46){\makebox(0,0)[lb]{\large $t_i-i\beta$}} \thicklines
\put(25,69){\circle*{1}} \put( 6,69){\circle*{1}}
\put(6,67){\circle*{1}} \put( 6,50){\circle*{1}}
\put(47,68){\circle*{1}} \put(46,68){\oval(2,2)[r]}
{
\put(32,70){\makebox(0,0)[lb]{\large    $C_{1}$}} \put(
6,69){\vector( 1, 0){28}} \put(34,69){\line( 1, 0){12}}

\put(18,62){\makebox(0,0)[lb]{\large    $C_{2}$}}
\put(46,67){\vector(-1, 0){28}} \put(18,67){\line(-1, 0){12}} }

{%
\put( 7,56){\makebox(0,0)[lb]{    {\large $C_3$}}} \put(
6,67){\vector( 0,-1){11}} \put( 6,56){\line( 0,-1){ 6}} }

\put(4,50){\line( 1, 0){ 2}} \put(50,62){\makebox(0,0)[lb]{\large
${\rm Re}  (\tau)$}} \put(-5,68){\vector( 1, 0){60}}
\put(-2,76){\makebox(0,0)[lb]{\large ${\rm Im} (\tau)$}}
\put(0,48){\vector( 0, 1){28}}



\end{picture}
\end{center}

\subsection{Effective Potential}

The effective potential should be the lowest term in the
derivative expansion of the effective action, i.e.\ equal to the
effective action for constant fields
\begin{equation}
\Sefft = -i \phi_i^2 \sum_{a,b=1}^3 \int_{C_a} d\tau \int d\vecx
\; B_{ab}
\end{equation}
The identities $B_{11}+B_{12} + B_{21} + B_{22} = 0$ and
$B_{13}+B_{23} = 0$ then tell us that in this form only
integrations along $C_3$ contribute to the total effective
potential. Another way of seeing this is to take $t_f=t_i$ leaving
$C= C_3$, $C_H = \emptyset$.  Thus
\begin{equation}
\Vteff \propto \phi_i^2 \int_{C_3} d\tau \int d \vecx \;  B_{33}
\end{equation}

However, I can get an alternative expression which is more
relevant to my analysis of the effective action.  Note\cite{TSEze}
that in equilibrium $B_{ab}$ is independent of $\tau$ by time
translation invariance. The integral over $\tau$ merely gives a
factor of $\beta$ to the overall volume factor out front. Thus
$\partial_\mu \phi = 0 \Rightarrow$
\begin{eqnarray}\label{veff}
\Steff &=&  - \half (-i\beta) V
  \phi_i^2 \sum_{b=1}^{3}  B_{ab} (\tau,\vecp=0,\Edd=0)
  \label{Seffconst}
  \nnel
  && \mbox{            } \forall \tau \in C_a, \; \; a=1,2,3
\end{eqnarray}
Without loss of generality choose $a=1, \tau \in C_1$ which gives
\begin{eqnarray}\label{veff2}
 \Vteff (\phi) &\propto &
 \phi_i^2\left[ ( B_{11} + B_{12})+ B_{13}\right]
\\
&=& \phi_i^2\left[  \Bret + B_{13}\right]
\end{eqnarray}
Note there are no ambiguities in the effective potential coming
from the $\Edd,\vecp \rightarrow 0$ limit of $B$ with whatever
$t_i \neq -\infty$ is used, so the lack of analyticity problem is
in this context again solved.  However it is {\em essential} to
have both horizontal $B_{11},B_{12}$ and vertical $B_{13}$
contributions to get correct answer. In $t_i \rightarrow -\infty$
both  horizontal $B_{11},B_{12}$ and vertical $B_{13}$ may be
needed but great care is needed with regulators to analyse this
limit, and this accounts for much of the confusion with real-time
calculations of the effective potential where just $\Bret$ is
often encountered.

\end{document}